\documentclass[journal=jctcce,manuscript=article,layout=twocolumn]{achemso}

\usepackage[colorlinks=true,citecolor=blue,linkcolor=blue,urlcolor=blue]{hyperref}
\usepackage{amsmath, amssymb}
\usepackage{graphicx, nicefrac, textcomp}
\usepackage{mwe}
\usepackage{braket}
\usepackage{color}
\usepackage{dcolumn}
\usepackage[utf8]{inputenc}
\usepackage{float}
\usepackage{dsfont}
\usepackage{svg}
\usepackage[version=4]{mhchem} % Formula subscripts using \ce{}

\usepackage[toc,page]{appendix}
\usepackage{soul}
\usepackage{algorithm}
\usepackage{algpseudocode}

 \usepackage[authormarkuptext=name,authormarkup=none,final]{changes}
\definecolor{red}{RGB}{255, 0, 0}
\definecolor{blue}{RGB}{0, 0, 255}
\definecolor{green}{RGB}{0, 192, 0}
\definechangesauthor[name={Auth.}, color={blue}]{GL}
\definechangesauthor[name={Auth.}, color={blue}]{PH}

\begin{tocentry}
   \includegraphics[width=\textwidth]{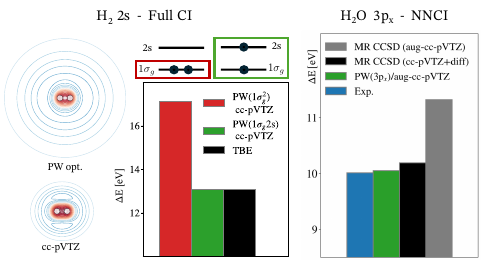}
\end{tocentry}

\title{Orbital Optimization and Neural-Network-Assisted Configuration Interaction Calculations of Rydberg States}
\author{Gianluca Levi}
    \affiliation{Science Institute and Faculty of Physical Sciences, University of Iceland, Reykjav\'ik, Iceland}
    \alsoaffiliation{Department of Chemical and Pharmaceutical Sciences, University of Trieste, 34127 Trieste, Italy}
    \email{gianluca.levi@units.it}

\author{Max Kroesbergen}
 \affiliation{Department of Physics, Friedrich-Alexander-Universit\"at Erlangen/N\"urnberg, 91058 Erlangen, Germany}
\author{Louis Thirion}
    \affiliation{Department of Physics, Friedrich-Alexander-Universit\"at Erlangen/N\"urnberg, 91058 Erlangen, Germany}
    \alsoaffiliation{Science Institute and Faculty of Physical Sciences, University of Iceland, Reykjav\'ik, Iceland}

\author{Yorick L. A. Schmerwitz}
    \affiliation{Science Institute and Faculty of Physical Sciences, University of Iceland, Reykjav\'ik, Iceland}
    \alsoaffiliation{Max-Planck-Institut f\"ur Kohlenforschung, 45470 M\"ulheim an der Ruhr, Germany}

\author{Elvar~\"O.~J\'onsson}
    \affiliation{Science Institute and Faculty of Physical Sciences, University of Iceland, Reykjav\'ik, Iceland}
\author{Pavlo Bilous}
    \affiliation{Max Planck Institute for the Science of Light, Staudtstraße 2, 91058 Erlangen, Germany}
\author{Philipp Hansmann}
    \affiliation{Department of Physics, Friedrich-Alexander-Universit\"at Erlangen/N\"urnberg, 91058 Erlangen, Germany}
    \alsoaffiliation{Science Institute and Faculty of Physical Sciences, University of Iceland, Reykjav\'ik, Iceland}
    \email{philipp.hansmann@fau.de}
\author{Hannes J\'onsson}
    \affiliation{Science Institute and Faculty of Physical Sciences, University of Iceland, Reykjav\'ik, Iceland}
    \email{hj@hi.is}

\begin{document}
%
% \date{\today}

\begin{abstract}
Rydberg excited states of molecules pose a challenge for electronic structure calculations because of their highly diffuse electron distribution. Even large and elaborate atomic basis sets tend to underrepresent the long-range tail, overly confining the Rydberg state.
An approach is presented 
% HJ add
here where the molecular orbitals are variationally optimized for the excited state using a plane wave basis set in
% HJ change to singular
a Hartree-Fock calculation, followed by 
% HJ add
a configuration interaction calculation. The use of excited state optimized orbitals greatly enhances the convergence of the many-body calculation, as illustrated by a full configuration interaction calculation of the $2s$ Rydberg state of \ce{H2}. A neural-network-based selective configuration interaction approach is then applied to calculations of 
%the 
$3s$ 
and 
$3p$
%, $3p_x$ and $3p_y$ 
states of \ce{H2O} and 
%the $3s$ and $3p_z$ states of 
\ce{NH3}. The obtained values of excitation energy are in close agreement with experimental measurements as well as previous many-body calculations 
where
%based on 
sufficiently diffuse atomic basis sets
were used.
Calculations using atomic basis sets lacking extra diffuse functions, such as aug-cc-pVTZ, give significantly higher estimates due to confinement of the Rydberg states.
\end{abstract}

\maketitle

\section{Introduction}
The accurate description of Rydberg excited states of molecules is a challenge in quantum chemistry because of the highly diffuse electron distribution. The energy of a set of Rydberg states of a molecule follows roughly the Rydberg series analogous to that of the hydrogen atom because the excited electron is subject to an effective potential consistent with an unscreened charge of nearly +e. 
For several molecules, such as \ce{NH3} and \ce{H2O}, even the first excited state is a Rydberg state. While time-dependent density functional theory (TDDFT) and equation-of-motion coupled cluster (EOM-CC) methods are widely used in calculations of excited states, they often struggle with calculations of Rydberg states due to limitations in the atomic orbital basis and approximations in the treatment of electron correlation \cite{Seidu2015, Cheng2008}. 
Accurate calculations of Rydberg states require special basis sets where extra diffuse functions have been added, as well as multi-reference representation of the wave function \cite{Mountaineering2021, Mountaineering2018, H2O_cc_2006}.

The application of Kohn-Sham DFT to Rydberg states is problematic because of the incorrect long-range form of the effective potential resulting from the self-interaction in the estimation of the classical electron-electron interaction from the total electron density. Recently, however, it has been demonstrated that orbital-optimized calculations using density functionals with explicit self-interaction corrections (SIC) can give remarkably accurate description of molecular Rydberg states \cite{OrbitaloptimizedSIC}. There, the convergence on excited states is obtained by converging to a saddle point on the electronic energy surface \cite{Schmerwitz23, Ivanov2021b, Levi2020}.
Furthermore, by using a real-space grid or plane wave representation, the diffuse nature of the Rydberg orbital can be represented better than with conventional diffuse atomic basis sets.
While such calculations provide an improved mean-field description, a more rigorous many-body treatment is required to accurately capture electron correlation effects in excited states and provide benchmarks for calibrating less accurate and faster methods.

In a recent study, we presented a neural-network (NN)-based selective configuration interaction (CI) approach to calculations of many-body wave functions for molecules \cite{N2_JCTC}.
The NNCI approach is based on an iterative classification of Slater determinants as relevant or irrelevant for the convergence of a chosen observable. 
In calculations of the ground state energy of molecules, it was found that NNCI is able to capture the correlation energy obtained by full CI results with several orders of magnitude fewer determinants. 
One of the key insights from this work is that an increase in the number of molecular orbitals (MOs) beyond what can be included in a typical full CI calculation improves significantly the accuracy of the results.

Building on these findings, the NNCI approach is extended here to calculations of excited electronic states, and applied to Rydberg states of molecules. A state-specific strategy is adopted, where the MOs are variationally optimized in plane wave-based Hartree-Fock calculations, and the resulting orbitals used in an NNCI calculation of the target excited state. 
The use of state-specific optimized orbitals has previously been shown to improve the results of CI calculations of excited states that account for single and double excitations (CISD)
%HJ change
%and using basis sets of 
with a basis of
localized atomic orbitals\cite{Kossoski2022, Decleva1994}. Here, a plane wave orbital representation is used in the Hartree-Fock calculations, ensuring full flexibility and an adequate description of the highly diffuse Rydberg states. Moreover, the CI calculations are not limited to single and double excitations, but the full Hilbert space is explored within 
%the 
a
selective NNCI approach.

The advantage of 
%using 
plane wave based Hartree-Fock orbitals optimized for the excited state is first illustrated in full CI calculations of the $2s$ Rydberg state of the \ce{H2} molecule. 
Then, the excited state NNCI approach is applied to Rydberg excitations of the \ce{NH3} and \ce{H2O} molecules, especially those that have previously turned out to be challenging for 
% HJ change
%typical 
traditional
quantum chemistry calculations \cite{OrbitaloptimizedSIC,Mountaineering2021,Mountaineering2018,H2O_cc_2006}.

% ------------------------------------------------------------------------------------
%
\section{Methods}
The Rydberg excited state calculations are performed by first converging either the ground or a target excited state in a Hartree–Fock calculation using a plane wave 
% HJ change
%basis to represent 
representation of
the occupied orbitals. The resulting Hartree–Fock MOs are then used 
%as the basis for a 
in subsequent CI calculations. 
\added[id=GL, comment=R2C2]{For each MO set, 
%we perform 
a full CI or NNCI calculation 
is carried out
% HJ remove
%in that same single-particle basis 
%and 
to obtain both the ground state and the target excited state as eigenstates of the same full CI/NNCI Hamiltonian matrix. The excitation energy is then calculated as the difference between the corresponding eigenvalues. This ensures orthogonality between the resulting CI/NNCI states within that MO basis.
% HJ add
The excitation energy obtained with the ground state MOs is compared with that obtained with the excited state MOs.
}
For the \ce{H2} molecule, a full CI calculation is performed, 
% HJ change
while the NNCI method is applied in a state-specific manner targeting the desired excited state
for the \ce{NH3} and \ce{H2O} molecules.
%the NNCI method is applied in a state-specific manner targeting the desired excited state. 
All calculations are performed by retaining spin symmetry.

% ---------------------   figure 1 ---------------------------
\begin{figure*}[t]
\includegraphics[width=1.7\columnwidth]{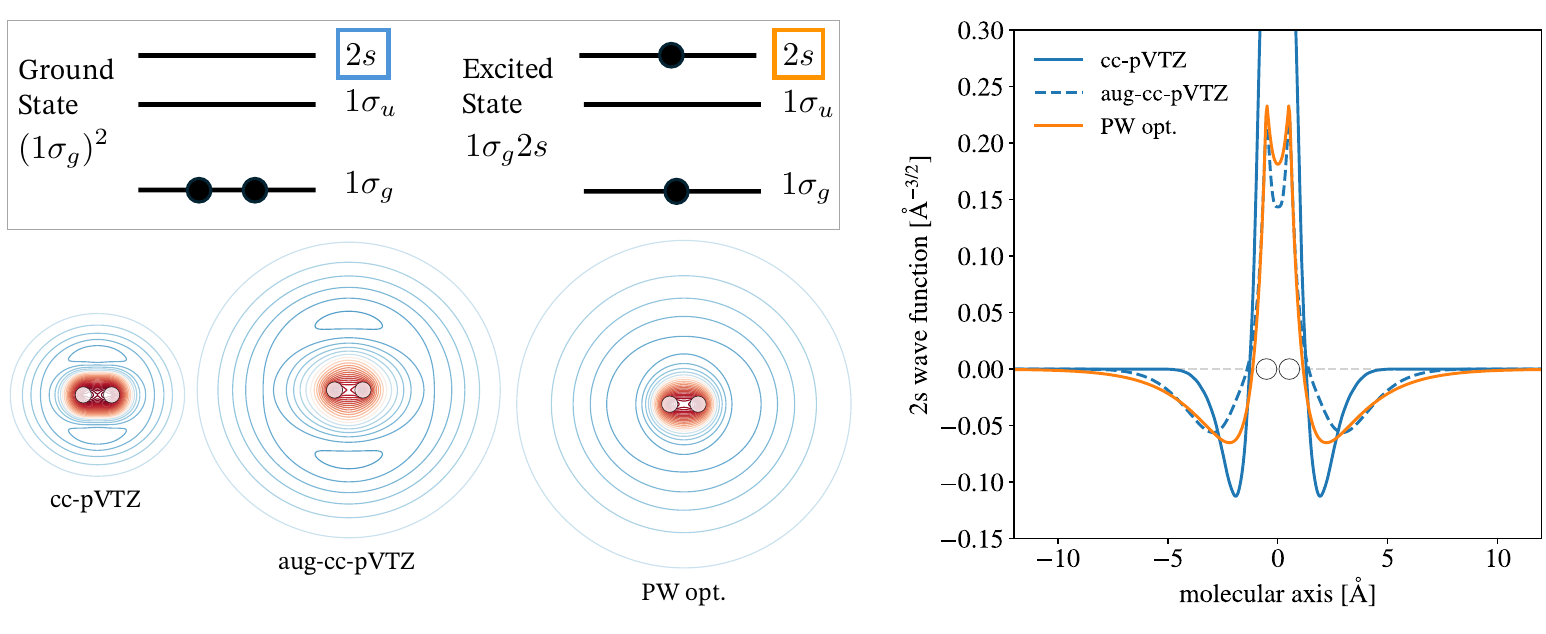}
\caption{
\label{fig:fig1} 
Top left: Orbital occupation in the spin restricted Hartree-Fock calculation of the ground and $2s$ Rydberg excited states of the \ce{H2} molecule. 
Bottom left and right: Comparison of the $2s$ Rydberg orbital obtained in three different calculations, displayed as contour maps (red and blue lines corresponding to regions of positive and negative amplitude, respectively) and amplitude along the molecular axis.
In a ground state Hartree-Fock calculation initialized using the cc-pVTZ (blue solid line) or aug-cc-pVTZ (blue dashed line) basis set, the $2s$ orbital has shorter tails than in an excited state calculation where the occupied orbitals are variationally optimized for the $2s$ Rydberg state (orange curve). In each case, the occupied orbitals are represented by a plane wave basis, but the unoccupied orbitals are largely determined by the atomic basis set used in the initialization.
Circles mark the positions of the protons.
}
\end{figure*}

% ------------------------------------------------------------
\subsection{Hartree-Fock Excited States}
The Hartree-Fock calculations use the projector augmented wave (PAW) method \cite{paw1, paw2} to treat the electrons near the nuclei, with core electrons frozen to the states obtained in scalar relativistic calculations of the isolated atoms. The MOs for the valence electrons are initialized using a localized atomic orbital basis of primitive Gaussian functions from the cc-pVTZ or aug-cc-pVTZ sets \cite{Dunning1989, Kendall1992}, with an additional set of numerical $s$-type atomic orbitals \cite{Rossi2015, Larsen2009}. Occupied orbitals are then variationally optimized in a plane wave representation, while unoccupied orbitals are only modified through the orthogonalization to the optimized occupied orbitals. 

The Hartree-Fock ground singlet state is obtained by minimizing the energy in a spin-restricted calculation using an L-BFGS algorithm \cite{Ivanov2021a}, thereby converging on a solution with aufbau orbital occupation. For the excited state Hartree-Fock calculations, the orbitals are initialized using the ground state orbitals setting the occupation numbers of the HOMO and the target excited orbital to 1. 
This corresponds to promotion of an electron from the HOMO to the target excited orbital within the spin-restricted representation. 
\added[id=GL, comment=R2C3]{The new set of restricted occupied orbitals, which includes the target Rydberg open-shell orbital, is variationally optimized in the plane wave basis set. In these excited state HF calculations, the energy is considered as a function of the one-electron reduced density matrix only, meaning that the explicit interaction between determinants of the open-shell singlet configuration state function is neglected\cite{Cernatic2022}.}%{While this orbital occupation does not correspond to a pure excited state, it ensures that the Rydberg orbital is variationally optimized in the plane wave basis set. The occupied orbitals are optimized, converging either to a local minimum (for the HOMO-LUMO excitation) or to a saddle point.} 
% HJ change
%The saddle 
The variational optimization of the orbitals for the excited state is carried out using a saddle point search algorithm
%point optimization is performed 
based on 
%using 
a direct optimization method \cite{Ivanov2021b} employing a limited-memory version of the symmetric rank-one (L-SR1) quasi-Newton algorithm \cite{Levi2020}. 

The Hartree-Fock calculations make use of a grid spacing of 0.18\,\text{\AA} and a plane wave energy cutoff of 1000\,eV. The simulation cell includes at least 10\,\text{\AA}  of vacuum between each atom and the nearest cell boundary. The calculations are performed with a development version of the GPAW software \cite{GPAW2024}. The atomic coordinates of the molecules are taken from Ref.\ \citenum{Haynes16}.

% ----
\subsection{NN-assisted CI}

For the calculation of the many-body wave functions, the Hamiltonian is cast into the form
\begin{equation}
\label{eq:Hfull}
H=H^0 +H^\mathrm{int} - \mathrm{MF}\left[H^\mathrm{int}\right]
\end{equation}
where 
\begin{align}
H^{0} &= \sum_{i,j,\sigma} t_{ij} c^\dagger_{i,\sigma} c_{j,\sigma},\\
H^\mathrm{int} &= \dfrac{1}{2}\sum_{\substack{i,j,k,l \\ \sigma,\sigma'}} U_{ijkl} \;\;c^\dagger_{i,\sigma} c_{j,\sigma}  c^\dagger_{k,\sigma'} c_{l,\sigma'}.
\end{align}
Here, $c^\dagger_{i,\sigma}$ and $c_{i,\sigma}$ are fermionic (creation/annihilation) field operators with orbital and spin indices $i$ and $\sigma$, respectively. The $t_{ij}$ and $U_{ijkl}$ are the single- and two-particle integrals
\begin{align}
t_{ij} &\equiv \int \text{d}\mathbf{r} \varphi^*_i(\mathbf{r}) \left(-\frac{1}{2}\nabla^2 + V^\text{eff}(\mathbf{r}) \right) \varphi_j(\mathbf{r})\\
U_{ijkl} &\equiv \int \text{d}\mathbf{r}\,\text{d}\mathbf{r'} \varphi^*_i(\mathbf{r}) \varphi_j(\mathbf{r}) \frac{1}{|\mathbf{r}-\mathbf{r}'|}  \varphi_k^*(\mathbf{r}') \varphi_l(\mathbf{r}')
\end{align}
where $V_\text{eff}(\mathbf{r})$ is the self-consistent mean-field potential, and the integrals are evaluated using the excited state optimized Hartree-Fock orbitals $\varphi_i(\mathbf{r})$. 
The term $\mathrm{MF}\left[H^\mathrm{int}\right]$ in Eq.~\eqref{eq:Hfull}, is the mean-field decoupled interaction operator. Since its contributions are implicitly included in the Hartree-Fock single particle integrals, $t_{ij}$, it needs to be subtracted in order to avoid double counting.
%

% HJ change
%All t
The CI and NNCI computations are performed using the {\sc SOLAX} package~\cite{SOLAX}, a Python library designed to compute and analyze fermionic quantum systems using the formalism of second quantization with built-in machine learning support,
with orbitals and matrix elements obtained in the Hartree-Fock calculations using the GPAW software.\\

\textbf{Selective CI - } The exact many-body eigenstates, $|\Psi_\alpha\rangle$, fulfill the eigenvalue equation for Hamiltonian \eqref{eq:Hfull} 
\begin{equation}
    H |\Psi_\alpha\rangle = E_\alpha |\Psi_\alpha\rangle
\end{equation}
and can be expanded in Slater determinants, $|\phi\rangle$ which form an orthonormal basis $\{ \phi\}$ of the full Hilbert space $\mathcal{H}^{\text{full}} = \text{span}(\{\phi\})$:
\begin{equation}
    |\Psi_\alpha\rangle = \sum_{\mathcal{H}^{\text{full}}} c^\alpha_{\phi} |\phi\rangle
\end{equation}
In selective CI, this is approximated by working within a subspace
$\mathcal{H}_s \subset \mathcal{H}_{\text{full}}$ with $\dim(\mathcal{H}_s) \ll \dim(\mathcal{H}_{\text{full}})$, 
such that
\begin{equation}
|\Psi_\alpha\rangle \approx |\Psi^\text{approx}_\alpha\rangle = \sum_{\phi \in \mathcal{H}_s} c^\alpha_\phi |\phi\rangle
\end{equation}

Our iterative protocol for the selective CI procedure starts from a small initial set of determinants, $\{\phi_{\text{init}}\}$. The set is then extended by an extension operator $\hat{\mathcal{O}}$ that projects out of the current subspace of the total Hilbert space and generates new determinants (provided that the set does not span an eigenspace of the operator). In each iteration, the Hamiltonian eigenproblem is solved within the corresponding subspace, yielding an approximate eigenstate wave function and total energy whose accuracy improves over successive iterations.
%due to the variational principle.
%
Due to the combinatorial growth in the number of generated configurations, inclusion of all new determinants quickly becomes impractical. Consequently, a selection protocol has to be introduced. 

\added[id=PH, comment={R1C4, R2C1,R1C2}]{Here, we follow Refs. \citenum{N2_JCTC} and \citenum{Bilous2024} for an NN-based selection scheme (NNCI) with modifications for initialization and state-tracking:
%
%Our previous NNCI calculations of ground electronic states demonstrate that the NN-based selection significantly outperforms simpler truncation schemes, such as the use of an energy cutoff.\\
%
%\textbf{NNCI Iterations - }
In the initialization, we define the set $\{\phi_{\text{init}}\}$ by performing an FCI calculation on a subset of (here ten) orbitals with the lowest energy. We find that - in practice - this strategy accelerates convergence compared to starting from only the Hartree Fock determinant. After this initialization, the iterative loop is started on the full orbital basis until convergence is reached.
In order to track the Rydberg excited state, we identify the target many-body (singlet) state $\Psi_\alpha$ with maximal occupation of the highly diffuse Rydberg orbital $n_\text{Ryd}^{\alpha}=\langle \Psi_\alpha|\hat n_\text{Ryd}|\Psi_\alpha\rangle$.
}

\begin{algorithm}[htb]
\caption{NNCI Iteration}\label{al:NNCI_iteration}
\begin{algorithmic}[1]
\State \textbf{Input:}
\begin{itemize}
    %\item Target state $|\Psi_\alpha\rangle$ to be converged 
    \item Determinants $\{\phi\}$ currently in CI expansion
    \item Pool of candidate determinants $\{\phi_{\text{pool}}\} = \mathcal{O}\{\hat{\phi}\} \setminus \{\phi\}$
    \item Target selection size %Number of most important to include from the pool
    \item Size of the random sample
\end{itemize}
\State Generate random sample $\{\phi_{\text{rand}}\} \subset \{\phi_{\text{pool}}\}$
\State Identify target state  $\alpha$ on $\{\phi\} \cup \{\phi_{\text{rand}}\}$
\vspace{0.1cm}
\State Adjust the cutoff parameter to split $\{\phi_{\text{rand}}\}$ into important $\{\phi_{\text{rand}}^{\text{impt}}\}$ and unimportant $\{\phi_{\text{rand}}^{\text{unimpt}}\}$ classes
\vspace{0.1cm}
\State Train NN classifier on $\{\phi\} \cup \{\phi_{\text{rand}}\}$
\vspace{0.1cm}
\State Predict important determinants \newline $\{\phi_{\text{NN}}^{\text{impt}}\} \subset \{\phi_{\text{pool}}\}$ %using trained NN
\vspace{0.1cm}
\State Identify target state $\alpha$
in \newline $\{\phi\} \cup \{\phi_{\text{NN}}^{\text{impt}}\} \cup \{\phi_{\text{rand}}^{\text{impt}}\}$
\vspace{0.1cm}
\State Eliminate false positives from $\{\phi_{\text{NN}}^{\text{impt}}\}$
\vspace{0.1cm}
\State Identify target state $\alpha$ in \newline $\{\phi\} \cup \{\phi_{\text{NN}}^{\text{impt}}\} \cup \{\phi_{\text{rand}}^{\text{impt}}\}$
\vspace{0.1cm}
\State \textbf{Output:} \newline $\alpha, \{\phi\} \leftarrow \{\phi\} \cup \{\phi_{\text{NN}}^{\text{impt}}\} \cup \{\phi_{\text{rand}}^{\text{impt}}\}$
\end{algorithmic}
\end{algorithm}
% ---------------------------------------------------------------------------------

In Algorithm~\ref{al:NNCI_iteration}, the steps following the initialization phase are summarized.
Note that the NNCI calculations are performed separately for the ground and the excited state for a given set of MOs.

% ----------------------------------------------------------------------
\section{Results and Discussion}
\subsection{Full CI for $2s$ state of \ce{H2}}
The advantage of optimizing the orbitals for the excited state in the Hartree-Fock calculation is first illustrated in calculations of the $2s$ Rydberg state of the \ce{H2} molecule. Here, 
%the 
full CI 
%basis is tractable directly 
calculations
%can easily be 
are
carried out
%and 
with
no NN-based selection.
%is needed.
%
Fig.~\ref{fig:fig1} shows the occupation of the orbitals in the ground and $2s$ Rydberg state as well as the shape of the $2s$ orbital obtained in three different calculations. 
In a calculation of the ground state, $(1\sigma_g)^2$, where the unoccupied $2s$ orbital is described within the cc-pVTZ basis set, the orbital is confined and lacks the long-range tails. 
With the extended aug-cc-pVTZ basis set, a significantly larger spread of the $2s$ orbital is obtained. 
But, in a calculation of the excited state, $1\sigma_g 2s$, where the occupied $2s$ orbital is represented by plane waves, the orbital becomes even more diffuse, reflecting more accurately its Rydberg character. 

\begin{figure}[H]
\includegraphics[width=0.9\columnwidth]{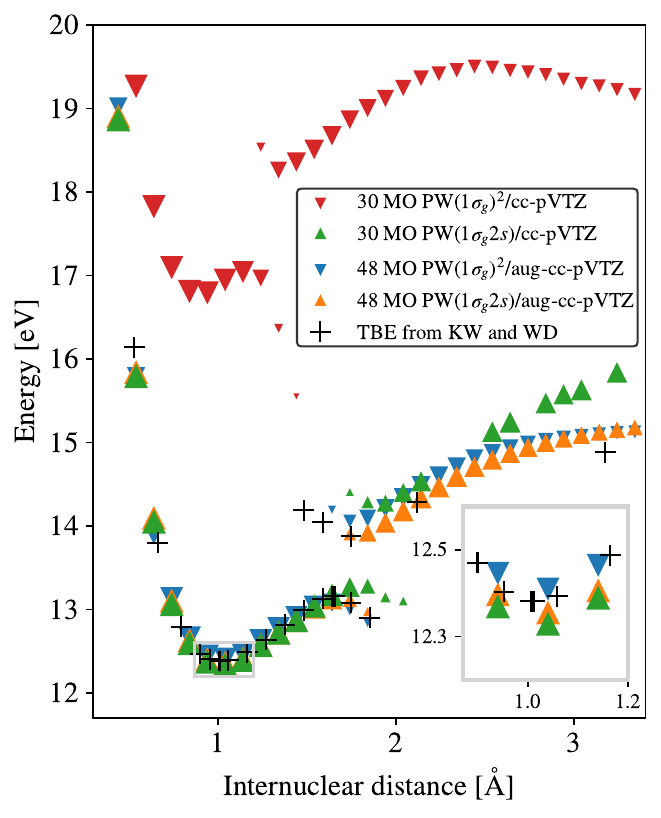}
\caption{
    \label{fig:H2_S3curve} 
    Energy of the $2s$ Rydberg state of \ce{H2} 
     with respect to the ground state energy minimum, 
    as a function of H-H distance obtained in full CI calculations using different sets of orbitals. The occupied orbitals are variationally optimized in Hartree-Fock plane wave (PW) calculations 
    % HJ change
    either
    for the ground ($(1\sigma_g)^2$, downward triangles), or for the $2s$ state ($1\sigma_g 2s$, upward triangles), while the unoccupied orbitals are represented with the cc-pVTZ or aug-cc-pVTZ atomic basis sets used in the initialization. The size of triangles reflects the overlap of the full CI many-body state with the $1\sigma_g 2s$ two Slater determinant wave function. Inset: Expanded scale around the minimum. Theoretical best estimates (TBE) are from Refs.\citenum{Kolos1969} (KW, lower branch) and \citenum{Wolniewicz1993} (WD, upper branch).
    }
\end{figure}

The energy of the $2s$ Rydberg state obtained in full CI calculations is illustrated in Fig.~\ref{fig:H2_S3curve} as a function of the internuclear separation and compared to the highly accurate theoretical best estimate (TBE) of Kolos, Wolniewicz and Dressler reported in Refs.\citenum{Kolos1969} and \citenum{Wolniewicz1993}. 

% ------------------------  figure 2 ------------------------
% --------------------------------------------------------------
%
The zero of energy is taken to be the minimum ground state energy. As the H-H bond is stretched, there is an avoided crossing with the $(1\sigma_u)^2$ state. 
The size of the triangles is determined by the overlap $\langle\Psi_{1\sigma_g2s}|\Psi_{\text{MB}}\rangle$, where $|\Psi_{\text{MB}}\rangle$ are the many-body eigenstates and
\begin{equation}
\Ket{\Psi_{1\sigma_g2s}}
= \frac{1}{\sqrt{2}}\Big(
\Ket{1\sigma_g\;\overline{2s}}
+\Ket{\overline{1\sigma_g}\;2s}
\Big).
\end{equation}
This projection indicates the weight of the  $1\sigma_g 2s$ contribution in each of the two branches near the avoided crossing.
All states are included in the graph, but states with a negligible overlap are not visible, as a small threshold is imposed. 

When the calculations are initialized with the cc-pVTZ basis set with 30 MOs, a large overestimation of the excitation energy is obtained if the occupied orbitals are optimized for the ground state, $(1\sigma_g)^2$.
At the optimal excited state bond length, the energy with respect to the minimum ground state energy is calculated to be $\sim$16.7\,eV, an overestimate by more than 4\,eV.
This is a result of the strong confinement of the Rydberg orbital as the $2s$ orbital obtained in the cc-pVTZ basis set lacks the long-range tails.
However, when the occupied orbitals are optimized for the $2s$ Rydberg state, $1\sigma_g 2s$, the calculated energy is close to the TBE, with only a slight underestimation of less than 0.1\,eV. Furthermore, the position of the avoided crossing with the $(1\sigma_u)^2$ state is then in good agreement with the TBE, while the calculation with orbitals optimized for the ground state shifts the avoided crossing to a shorter bond distance by about 0.5\,\text{\AA}.

When the aug-cc-pVTZ basis set with 48 MOs is used for the initialization of the Hartree-Fock ground state calculation, the unoccupied $2s$ Rydberg orbital has significantly longer tails and the calculated energy is improved compared to that obtained with the cc-pVTZ basis set, as can be seen in Fig.~\ref{fig:H2_S3curve}.
The excitation energy, however, remains slightly overestimated, as the tails of the $2s$ orbital are still not sufficiently extended (see Fig.~\ref{fig:fig1}). 
A further improvement is obtained when the orbitals are optimized for the $2s$ state instead of the ground state, particularly in the internuclear distance range of 1.5–3\,\text{\AA}. 
%Under this condition
Then, the energy gap 
%at the avoided crossing is reproduced with near-perfect accuracy.
of the TBE is reproduced almost perfectly,
as can be seen
%The results shown 
in Fig.~\ref{fig:H2_S3curve}.
%illustrate this. 

In the ground state calculation, the unoccupied orbitals, and thereby the $2s$ Rydberg orbital, are not optimized. 
Instead, they are represented with the cc-pVTZ or aug-cc-pVTZ basis sets and only get adjusted by the orthogonalization to the optimized orbitals. 
Therefore, the unoccupied $2s$ Rydberg orbital remains limited by the localized nature of the atomic orbitals basis set. 
In contrast, the plane wave based optimization of the $2s$ orbital in the excited state Hartree-Fock calculation describes the diffuse nature of the Rydberg state 
%much better 
and provides significantly improved results. 

For larger molecules, a full CI calculation becomes computationally demanding and the NNCI provides an efficient alternative by including only the most important determinants. 
In excited state calculations, the efficiency is further enhanced by optimizing the Hartree-Fock orbitals for the target state. 
This is illustrated here in calculations of rather low lying Rydberg states of the \ce{NH3} and \ce{H2O} molecules, and becomes especially clear already for the $3p_z$ state of \ce{NH3} and the $3p_x$ state of \ce{H2O}.

%   Ammonia results:
\subsection{NNCI for \ce{NH3} Rydberg States}

% --------------  figure 3 -----------
\begin{figure}[t]
    \includegraphics[width=0.9\columnwidth]{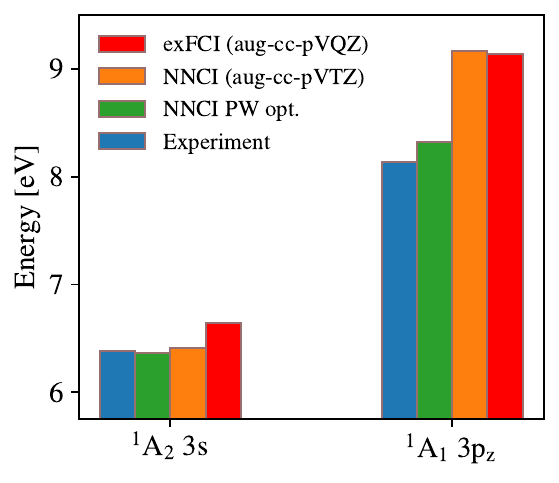}
    \caption{
    Excitation energy for $3s$ and $3p_z$ Rydberg states of the \ce{NH3} molecule calculated with NNCI using 52 MOs, formed initially from the aug-cc-pVTZ basis set, but then optimized variationally for the occupied ones in plane wave based Hartree-Fock calculations, either for the ground state (NNCI aug-cc-pVTZ) or for the target Rydberg state (NNCI PW opt). 
    For comparison, experimental estimates from Refs.\citenum{Skerbele65} and \citenum{Arfa91} are shown as well as extrapolated full CI calculations (exFCI) from Ref.\citenum{Mountaineering2018} based on the aug-cc-pVQZ basis set. 
    }
    \label{fig:NH3_barplot}
\end{figure}
% -------------------------------------
Fig.~\ref{fig:NH3_barplot} shows results of NNCI calculations for the $3s$ and $3p_z$ Rydberg states of \ce{NH3}. 
The $z$-axis is the C$_3$ rotational symmetry axis. A comparison with experimental measurements from Refs.\citenum{Skerbele65} and \citenum{Arfa91}, as well as reported extrapolated full CI (exFCI) calculations from Ref.\citenum{Mountaineering2018} is shown. 
For both $3s$ and $3p_z$ states, the NNCI calculated excitation energy is in close agreement with the measured values.

The exFCI calculations are based on the aug-cc-pVTZ basis and give a slightly higher excitation energy than NNCI for the $3s$ state and a significantly higher one for the $3p_z$ state, by about 0.8\,eV.
Even though the NNCI calculation is initialised with the same atomic basis set, 
a better description of the Rydberg state is obtained and fewer determinants are needed in the CI calculation to reach convergence
because the occupied orbitals are variationally optimized for the excited state in the Hartree-Fock calculation. 

The use of ground state orbitals in a CI calculation based on the aug-cc-pVTZ basis set overly confines the Rydberg state and thereby overestimates the excitation energy. 
This is verified by carrying out the NNCI calculation using the ground state orbitals, which indeed gives a higher estimate of the $3p_z$ excitation energy, similar to that of the exFCI calculations, as shown in Fig.~\ref{fig:NH3_barplot}.
Again the reason is the missing long-range tail of the $3p_z$ orbitals obtained from the ground state calculations with the aug-cc-pVTZ basis set. 

The limitation of the aug-cc-pVTZ basis set for the $3p_z$ state of \ce{NH3} has previously been demonstrated in variational density functional calculations where the atomic basis set calculations were compared to calculations carried out with a real space grid, see Figs. 2 and 3 in Ref.\citenum{OrbitaloptimizedSIC}.

% --------------  figure 4 ----------------
\begin{figure}
    \centering
    \includegraphics[width=0.9\columnwidth]{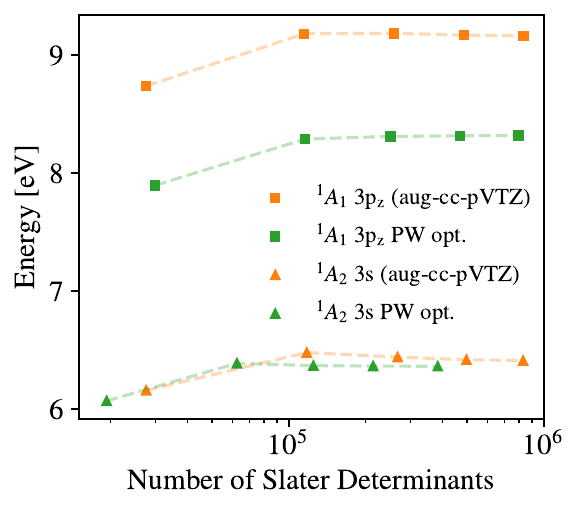}
    \caption{   
    Convergence of the excitation energy for the $3s$ and $3p_z$ Rydberg states of \ce{NH3} with respect to the number of Slater determinants included in the NNCI calculation, in all cases generated from 52 MOs formed initially from the aug-cc-pVTZ basis set. The occupied orbitals are optimized in plane wave based Hartree-Fock calculations either for the ground state (orange) or the target excited state (green).
    }
    \label{fig:NH3_convergence}
\end{figure}
% ------------------------------------------

The convergence of the excitation energy with respect to NN iterations is illustrated for both the $3s$ and $3p_z$ Rydberg states of \ce{NH3} in Fig.~\ref{fig:NH3_convergence}.  
In four cycles of the NN-assisted extension of the set of included determinants and retraining of the NN, convergence is reached already with 10$^5$ determinants, which is five orders of magnitude less than what a full CI calculation would require.     
%
%Different from the $3s$ Rydberg state, the energy of the $3p_z$ state is overestimated when the orbitals are optimized for the ground state.
%
Further information on the convergence with respect to both the number of orbitals and determinants included in the NNCI can be found in the Supporting Information (SI).

% -----------------------------------------------
\subsection{NNCI for \ce{H2O} Rydberg States}

% --------------  figure 5 ----------------
\begin{figure}[t!]
        \includegraphics[width=0.9\columnwidth]{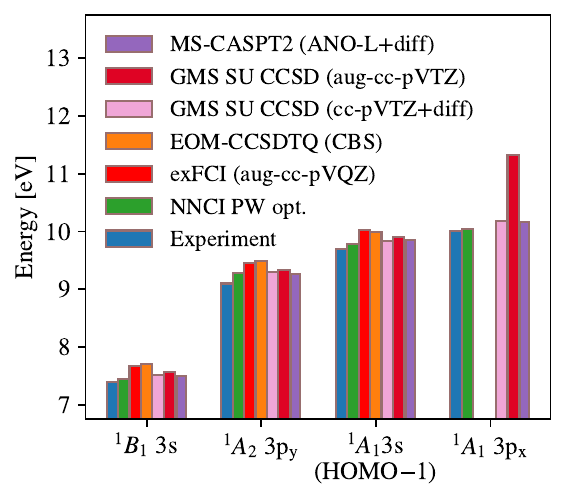}
    \caption{
    \label{fig:H2O_barplot} 
    Comparison of NNCI calculated excitation energy for the $3s$, $3p_y$\added[id=GL]{, $3s$ from HOMO-1}, and $3p_x$ Rydberg states of \ce{H2O}, based on 52 MOs initially formed from the aug-cc-pVTZ basis set 
    but then optimized variationally \added[id=GL]{for the occupied ones in plane wave based calculations} for the target Rydberg state (NNCI PW opt), with experimentally measured \cite{Chutjian75} values,
    illustrating good agreement.
    Results of several other theoretical calculations are also shown:
    extrapolated full CI calculations (exFCI) based on the aug-cc-pVQZ basis set \cite{Mountaineering2018},
    EOM-CCSDTQ calculations in the complete basis set (CBS) limit \cite{Mountaineering2021}, available only for three of the states, multireference GMS SU CCSD calculations \cite{H2O_cc_2006} based on either a cc-pVTZ+diff basis set that includes extra diffuse functions, or the standard aug-cc-pVTZ 
    basis set\added[id=GL]{, as well as multistate CASPT2 (MS-CASPT2) calculations based on a basis set of atomic natural orbitals supplemented with diffuse functions\cite{Rubio2008}}. \added[id=GL]{The multireference GMS SU CCSD calculations based on the aug-cc-pVTZ basis set}%{The latter} 
    overly confines the $3p_x$ leading to a large overestimate of the excitation energy, by more than 1\,eV.
    }
\end{figure}
% -------------------------------------

% --------------  figure 6 ----------------
\begin{figure}[b!]
    \includegraphics[width=0.9\columnwidth]{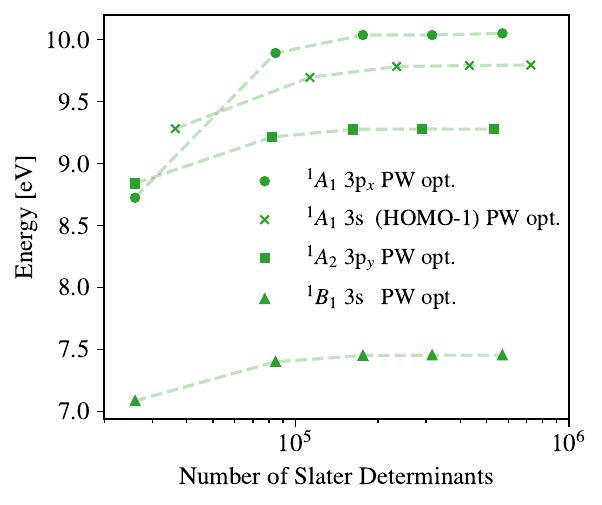}
    \caption{
        \label{fig:H2O_convergence}
        Convergence of the excitation energy for the $3s$, $3p_y$ and $3p_x$ from HOMO \added[id=GL]{as well as $3s$ from HOMO-1}   Rydberg states of \ce{H2O} with respect to the number of Slater determinants included in the NNCI calculations, in all cases generated from 52 MOs formed initially from the aug-cc-pVTZ basis set. The occupied orbitals are optimized in plane wave based calculations of the target excited state. 
    }
\end{figure}
% -------------------------------------

The results of the NNCI calculations for the \ce{H2O} molecule are shown in Fig.~\ref{fig:H2O_barplot}. The excitation energy obtained 
\added[id=GL]{for transitions from the HOMO to $3s$, $3p_y$ and $3p_x$ 
%states
and from HOMO-1 to $3s$\ \ }%{$3s$, $3p_x$ and $3p_y$ states} 
is shown and compared to experimental estimates \cite{Chutjian75} as well as 
results of
several previously reported
%high-level 
calculations. The $x$-axis is normal to the plane of the molecule. 
\added[id=GL, comment=R1C1]{The two lowest $^1A_1$ excited states are found to have 
%a 
strong multiconfigurational character, arising from the mixing of $3s$ from HOMO-1 and $3p_x$ from HOMO 
configurations. 
The NNCI calculations of these two states employ the same set of MOs, with the occupied orbitals optimized for the $3p_x$ configuration within a PW basis set. Since the $3p_x$ orbital is more diffuse than the $3s$ orbital, the latter is well described by the aug-cc-pVTZ atomic basis set used for the orbital initialization. In the NNCI calculations 
%employing 
with
52 MOs, the lower $^1A_1$ state is a mixture of the $3s$ from HOMO-1 and $3p_x$ from HOMO configurations with weights of $\sim$55\% and $\sim$40\%, respectively (see Fig. S8 in the SI), and is labelled ``$3s$ (HOMO-1)''. The upper $^1A_1$ state is characterized 
%by a mixing of the $3s$ from HOMO-1 and $3p_x$ configurations with 
by
weights of $\sim$40\% and $\sim$55\%, respectively, 
and is labelled ``$3p_x$''. 
The strong multiconfigurational character for these states has also been 
%obtained 
noted
in previous multi-state 
CASPT2 (MS-CASPT2) calculations using a basis set of atomic natural orbitals supplemented with diffuse functions centered on the oxygen atom \cite{Rubio2008}. The energy difference between the $3s$ from HOMO-1 and $3p_x$ from HOMO states predicted by the NNCI calculations with 52 MOs (0.26 eV, see Fig. S7 in the SI) agrees closely with the corresponding MS-CASPT2 value (0.29 eV).}

The agreement with the experimental measurements and with other theoretical values is good except for the GMS SU CCSD calculation \cite{H2O_cc_2006} of the $3p_x$ state when the latter is based on the aug-cc-pVTZ basis. 
This is a high-level multireference (9R) calculation, but it is limited by the atomic basis set and the calculated excitation energy is significantly larger than the value obtained with NNCI where the occupied orbitals are optimized for the $3p_x$ state. 
The difference is more than 1\,eV.
This is analogous to the results for the $3p_z$ state of the \ce{NH3} molecule shown in  Fig.~\ref{fig:NH3_barplot} and can be ascribed to confinement of the Rydberg state by the atomic basis set.
However, a GMS SU CCSD calculation with extra diffuse basis functions, cc-pVTZ+diff, gives a value of excitation energy close the results of the NNCI calculation. 
Again, this shows the advantage of variationally optimizing the orbitals for the target excited state in the PW-based Hartree-Fock calculations.
Even with the aug-cc-pVTZ basis for the initialization of the MOs, the results obtained with the NNCI calculations agree well with both experiment and high-level calculations with an extended atomic basis. 
EOM-CCSDTQ calculations in the complete basis set (CBS) limit from Ref.\citenum{Mountaineering2021} \added[id=GL]{and exFCI calculations with the aug-cc-pVQZ basis set from Ref.\citenum{Mountaineering2018}} are also in good agreement with the NNCI calculations of the \added[id=GL]{$3s$, $3p_y$, and $3s$ from HOMO-1}%{$3s$ and $3p_y$} 
 Rydberg states, but analogous calculations for the $3p_x$ Rydberg state are not available.

The limitation of the aug-cc-pVTZ basis for the $3p$ Rydberg states of \ce{H2O} has previously been demonstrated in the context of variational density functional calculations where a comparison was made with calculations using a real space grid \cite{OrbitaloptimizedSIC}.

Fig.~\ref{fig:H2O_convergence} illustrates the convergence of the NNCI calculation with respect to the number of NN cycles and the number of determinants included.
In four cycles, convergence is reached already with 10$^5$ determinants, similar to the \ce{NH3} calculation. 
Further details on the convergence are provided in the SI.

% -----------------------------------------------------------------------------
\section{Conclusion}

The results presented here show two important aspects of CI calculations of excited states. First of all, faster convergence 
% HJ change
in selective CI 
is obtained when the orbitals are variationally optimized for the target excited state in the Hartree-Fock calculations, here with a plane wave representation to overcome the limitations of typical atomic basis sets, such as aug-cc-pVTZ, when describing diffuse Rydberg states.
Secondly, full CI results for the excitation energy can be obtained with several orders of magnitude fewer Slater determinants by using the NN-based selective CI.
The NNCI method is extended here to calculations of excited electronic states and is shown to give accurate results for challenging Rydberg states of \ce{NH3} and \ce{H2O} molecules. 
Convergence is reached in four training/refinement NN iteration, leading to inclusion of 10$^5$ Slater determinants. 

The optimization of the orbitals for the target excited state demonstrated here to be important in calculations of Rydberg states, can also be of advantage for other types of excited states, for example long-range charge transfer excitations. 
There, the orbitals can differ significantly between the ground states and an optimized excited state and the number of determinants needed to reach convergence in a selective CI calculation is expected to be significantly reduced by using optimized orbitals. 

\added[id=GL, comment={R1C1, R1C4}]{The NNCI method has been applied here in calculations of both single-configurational and multiconfigurational excited states (the $3s$ from HOMO-1 and $3p_x$ states of the H$_2$O molecule). As other selective CI approaches, the NNCI method does not rely on the assumption of a single dominant configuration, and can describe multiconfigurational states by incorporating determinants that contribute with comparable weights. For multiconfigurational states, the orbitals participating to the dominant configurations need to be 
% HJ change
%all 
represented well to achieve high accuracy. For the $3s$ from HOMO-1 and $3p_x$ excited states of the H$_2$O molecule, only the most diffuse $3p_x$ orbital had to be optimized with a PW basis set, since the $3s$ orbital is more compact and well described by the aug-cc-pVTZ basis set. In other cases, e.g. when configurations with two equally diffuse orbitals mix, it might become important to optimize the molecular orbitals for a multideterminant reference instead of a single-determinant HF reference.}  

\added[id=PH, comment={R1C5}]{Finally we note, that the present framework based on a plane wave basis naturally suggests extensions to continuum and shape-resonance problems, where the target ``state'' is metastable and embedded in (or coupled to) a continuum.} %Such open quantum systems typically require effective non-Hermitian descriptions. A combination with excited state orbital optimization and NN-driven determinant selection would provide a concrete route to treat diffuse resonance wave functions beyond standard Gaussian basis sets, while the state-tracking strategy described above can be generalized to follow the targeted resonance by its wave function character rather than by energy ordering.}
% -----------------------------------------------------------------------------

\begin{acknowledgement} 
\added[id=GL]{The authors thank Lorenzo Restaino for fruitful discussions.} This work was supported by the Icelandic Research Fund (grants nos.\ 2511544 and 2410644) and the University of Iceland Research Fund. G.L. acknowledges support from the ERC under the European Union's Horizon Europe research and innovation programme (grant no. 101166044, project NEXUS). Views and opinions expressed are however those of the author(s) only and do not necessarily reflect those of the European Union or ERC Executive Agency. Neither the European Union nor the granting authority can be held responsible for them. P.B.\ gratefully acknowledges the ARTEMIS funding via the QuantERA program of the European Union provided by German Federal Ministry of Education and Research under the grant 13N16360 within the program ``From basic research to market''. Y.L.A.S.\ and P.B.\ acknowledge support by the Max Planck Society. Computer resources, data storage, and user support were provided by the Icelandic Research e-Infrastructure (IREI) funded by the Icelandic Infrastructure Fund. The authors further gratefully acknowledge the scientific support and HPC resources provided by the Erlangen National High Performance Computing Center (NHR@FAU) of the Friedrich-Alexander-Universität Erlangen-Nürnberg (FAU).   
\end{acknowledgement}

\section*{Supporting Information}
The convergence of the energy of the ground and excited states as well as the excitation energy as a function of the number of
Slater determinants and the number of molecular orbitals included in the NNCI calculations is illustrated in the Supporting Information. \added[id=GL]{A graph showing the weights of the $3s$ from HOMO-1 and $3p_x$ configurations for the NNCI many-body $3s$ from HOMO-1 and $3p_x$ states as a function of the number of molecular orbitals is also shown in the Supporting Information.}

\bibliography{main_references}

% -----------------------------------------------------------------------------

\end{document}

%
% \date{\today}

\maketitle

%\begin{suppinfo}
%\section*{Supporting Information}

The energy of the ground and excited state as well as the excitation energy as a function of the number of
Slater determinants and the number of molecular orbitals included in the NNCI calculations is illustrated in the following figures. 

\begin{figure*}[t]
\includegraphics[width=\columnwidth]{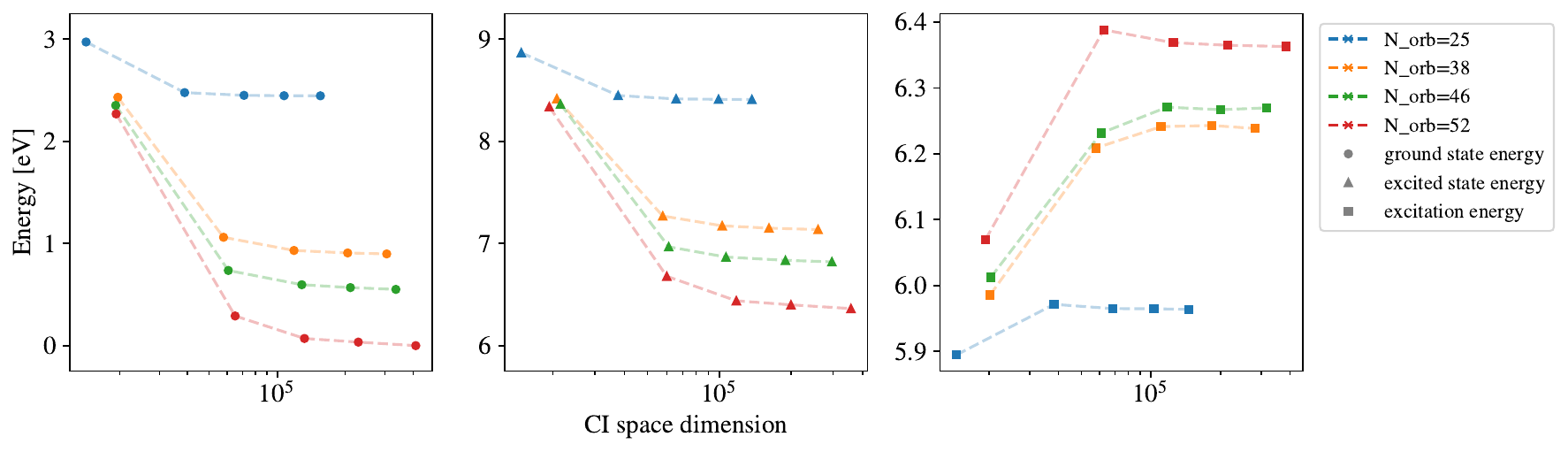}
\caption{\label{fig:NH3_s3} 
NNCI calculations to obtain the excitation energy of the $3s$ Rydberg state of the \ce{NH3} molecule 
using molecular orbitals optimized for the excited state. 
Left: ground state energy. 
Middle: excited state energy. 
Right: excitation energy. 
%All energies are given in eV.
}
\end{figure*}

\begin{figure*}[t]
\includegraphics[width=\columnwidth]{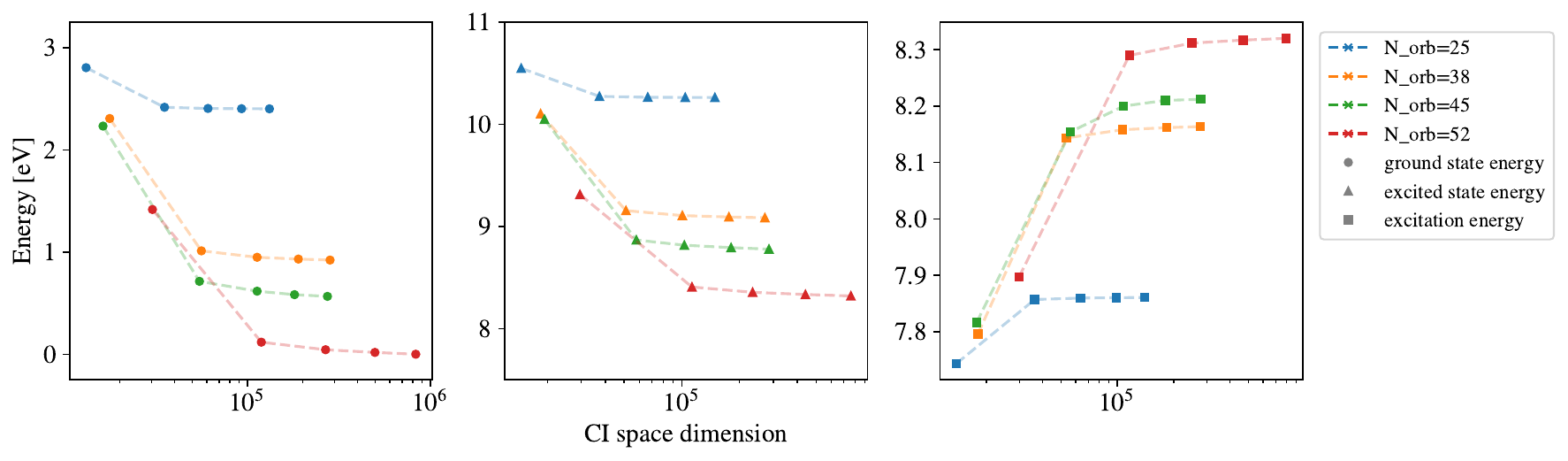}
\caption{\label{fig:NH3_R3pz} 
NNCI calculations to obtain the excitation energy of the $3p_z$ Rydberg state of the \ce{NH3} molecule 
using molecular orbitals optimized for the excited state. 
Left: ground state energy. 
Middle: excited state energy. 
Right: excitation energy. 
%All energies are given in eV.
}
\end{figure*}

\begin{figure*}[t]
\includegraphics[width=\columnwidth]{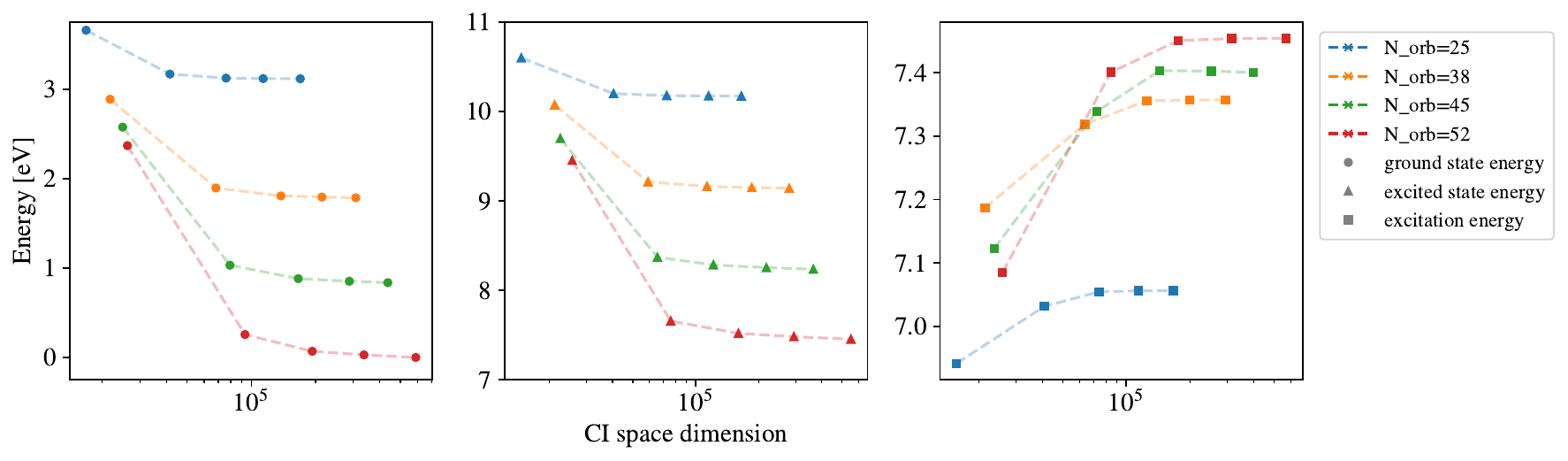}
\caption{\label{fig:H2O_0to0} 
NNCI calculations to obtain the excitation energy of the $3s$ Rydberg state of the \ce{H2O} molecule 
using molecular orbitals optimized for the excited state. 
Left: ground state energy. 
Middle: excited state energy. 
Right: excitation energy. 
%All energies are given in eV.
}
\end{figure*}

\begin{figure*}[t]
\includegraphics[width=\columnwidth]{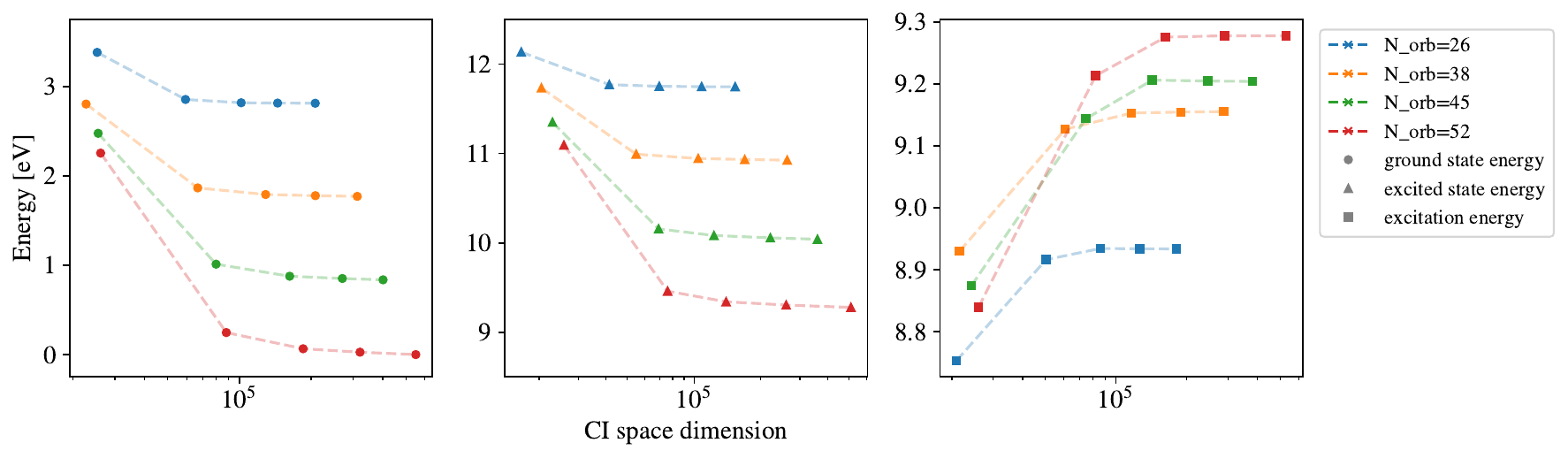}
\caption{\label{fig:H2O_0to1} 
NNCI calculations to obtain the excitation energy of the $3p_y$ Rydberg state of the \ce{H2O} molecule 
using molecular orbitals optimized for the excited state. 
Left: ground state energy. 
Middle: excited state energy. 
Right: excitation energy. 
%All energies are given in eV.
}
\end{figure*}

\begin{figure*}[t]
\includegraphics[width=\columnwidth]{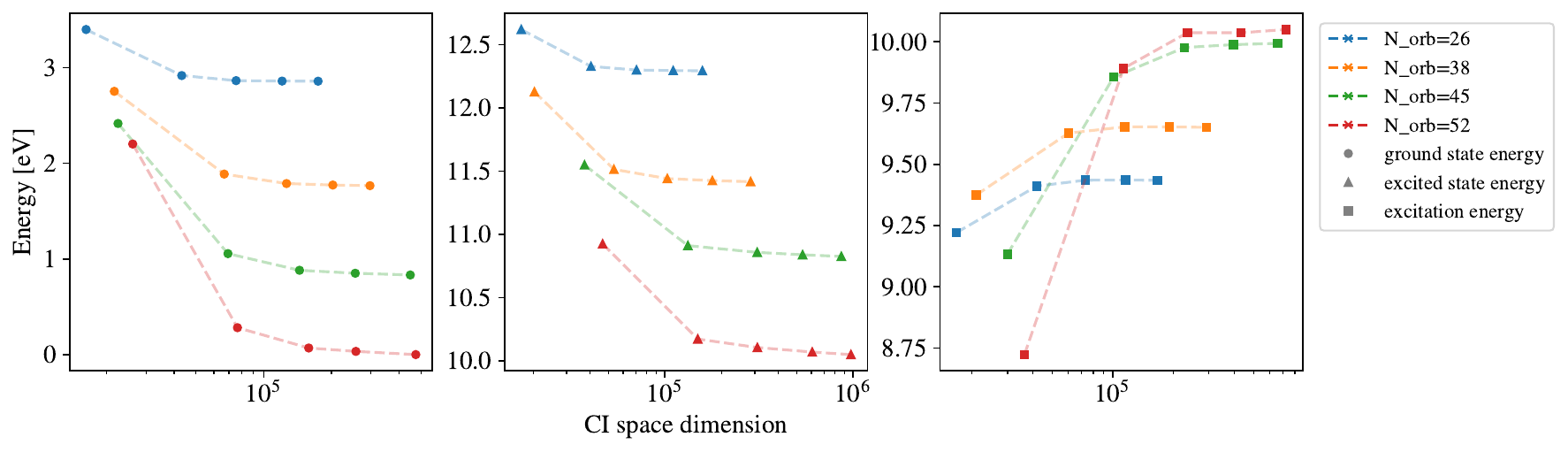}
\caption{\label{fig:H2O_0to3_px} 
NNCI calculations to obtain the excitation energy of the $3p_x$ Rydberg state of the \ce{H2O} molecule 
using molecular orbitals optimized for the excited state. 
Left: ground state energy. 
Middle: excited state energy. 
Right: excitation energy. 
%All energies are given in eV.
}
\end{figure*}
\begin{figure*}[t]
\includegraphics[width=\columnwidth]{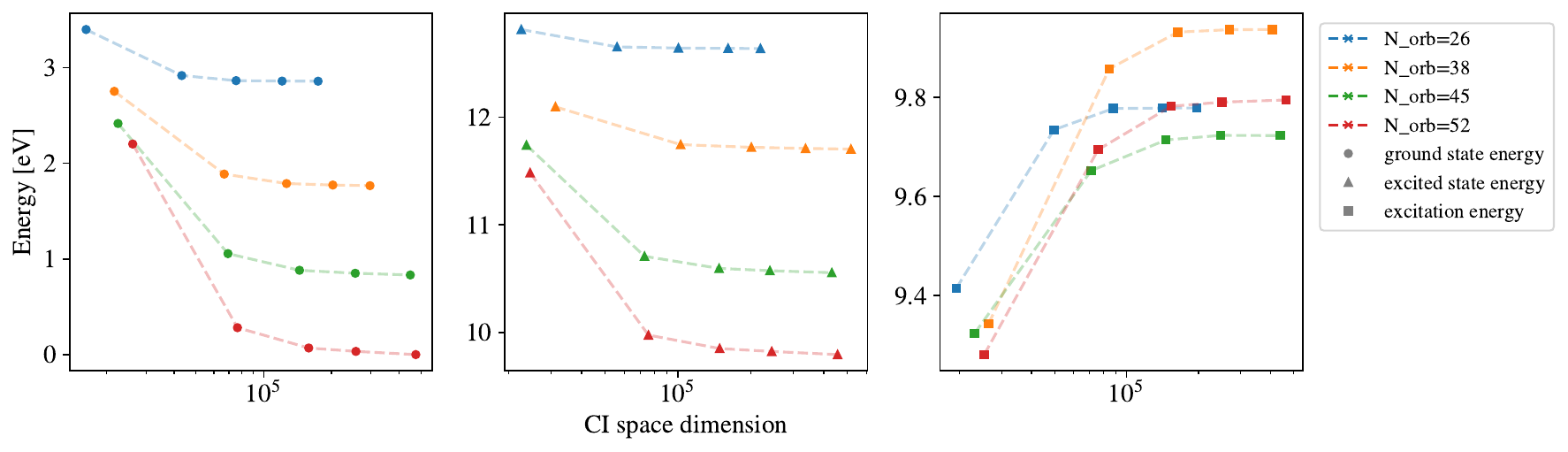}
\caption{\label{fig:H2O_0to3_3s} 
NNCI calculations to obtain the excitation energy of the $3s$ from HOMO-1 Rydberg state of the \ce{H2O} molecule 
using molecular orbitals optimized for the $3p_x$ 
excited state. 
Left: ground state energy. 
Middle: excited state energy. 
Right: excitation energy. 
%All energies are given in eV.
}
\end{figure*}

\begin{figure}[t]
\includegraphics[width=0.49\columnwidth]{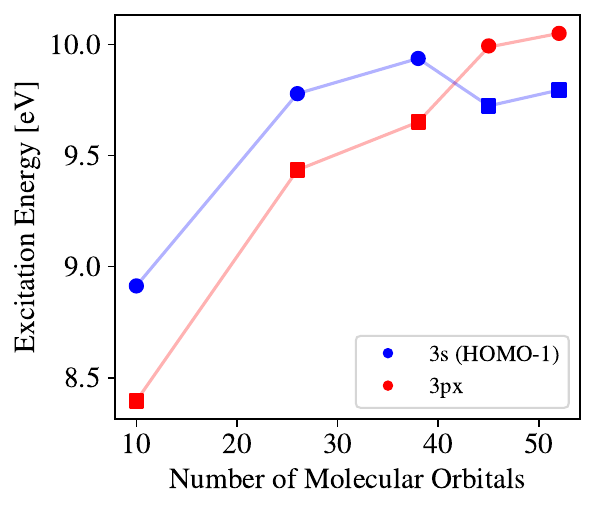}
\caption{\label{fig:H2O_0to3_px_3s_character} 
Excitation energy of the $3s$ from HOMO-1 and $3p_x$ from HOMO Rydberg states of the \ce{H2O} molecule using molecular orbitals (MOs) optimized for the $3p_x$ state. Both states have a multiconfigurational character, arising from mixing of the $3s$ from HOMO-1 and $3p_x$ excitated state configurations. For the state labelled ``$3s$ (HOMO-1)'', the $3s$ from HOMO-1 has larger weight, while for the state labelled $3p_x$, the $3p_x$ excitation dominates (see Fig. S8). As the number of MOs is increased, the ordering reverses, and for 52 MOs the $3p_x$ state is higher in energy than the $3s$ from HOMO-1 state by 0.26 eV.}
\end{figure}

\begin{figure}[t]
\includegraphics[width=0.49\columnwidth]{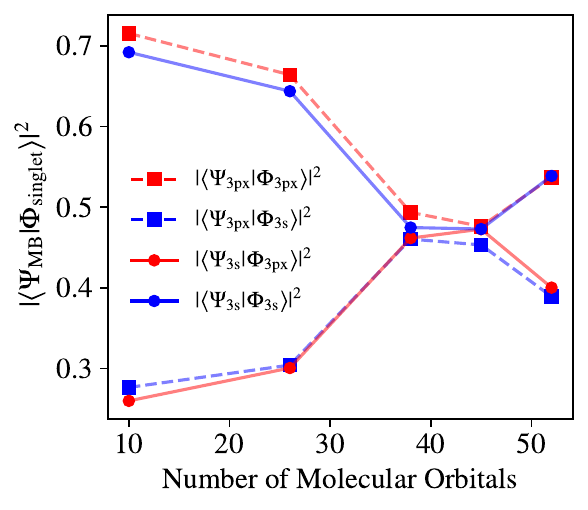}
\caption{\label{fig:H2O_0to3_px_3s_coeffs} 
Square of the overlap between the NNCI many-body states dominated by $3s$ from HOMO-1 ($|\psi_\text{3s}\rangle$) and $3p_x$ from HOMO ($|\psi_\text{3px}\rangle$) configurations with the corresponding \emph{pure} singlet states ($3s$ from HOMO-1: $|\phi_\text{3s}\rangle$ and $3p_x$ from HOMO: $|\phi_\text{3px}\rangle$). We show the overlap as a function of the number of molecular orbitals (optimized for the $3p_x$ Rydberg excitation) used in the NNCI calculation.%s, representing the weights of each singlet in the many-body states. 
%The molecular orbitals are those optimized in Hartree-Fock calculations for the $3p_x$ state.
}
\end{figure}

%\end{suppinfo}
%\appendix